\documentclass[aip,jcp,reprint]{revtex4-1}
\usepackage{graphicx}
\usepackage{rotating}
\usepackage{amsfonts,amsmath,latexsym,bbm,dsfont}

\usepackage{color}

\usepackage{tikz}
\newcommand{\nl}{\nonumber \\ && }

\newcommand{\beq}{\begin{eqnarray}}
\newcommand{\eeq}{\end{eqnarray}}
\newcommand{\half}{\frac{1}{2}}

\newcommand{\op}{\hat}
\newcommand{\mat}{\mathbf}

\newcommand{\eq}[1]{Eq.~(\ref{#1})}

\newcommand{\sect}[1]{sec.~\ref{#1}}
\newcommand{\fig}[1]{Figure~\ref{#1}}
\newcommand{\tab}[1]{Table~\ref{#1}}

\newcommand{\Perm}[2]{\hat{\cal P}^{#1}_{#2}}

\def\RCS$#1: #2 ${\@namedef{RCS#1}{#2}\typeout{RCS #1: #2}}
\mathchardef\lt="313C \mathchardef\gt="313E
\mathcode`<="4268 \mathcode`>="5269

\renewcommand\vec\mathbf

\begin{document}

\author{Daniel Kats}
\thanks{To whom correspondence should be addressed}
\affiliation{
Max Planck Institute for Solid State Research, Heisenbergstra\ss e 1, 70569 Stuttgart, Germany}
\author{Andreas K\"ohn}
\affiliation{Institut f\"ur Theoretische Chemie, Universit\"at Stuttgart, Pfaffenwaldring 55, 
D-70569 Stuttgart, Germany}

\title{On the distinguishable cluster approximation for triple excitations}
\begin{abstract}
The distinguishable cluster approximation applied to coupled cluster doubles equations 
greatly improves absolute and relative energies. 
We apply the same approximation to the triples equations and demonstrate that it can also 
improve results of the coupled cluster with singles, doubles and triples.
The resulting method has a nominal computational scaling of ${\cal O}(N^7)$ in the real-space representation, 
is orbital invariant, size extensive, and exact for three electrons.
\end{abstract}

\maketitle 
\section{Introduction}
The coupled cluster hierarchy of methods \cite{cizek:66} is known to converge very quickly to full configuration interaction (FCI)
results with comparably low excitation level for weakly correlated systems. 
Already triple excitation are usually sufficient to reach the 1 kcal/mol accuracy in relative energies,
\cite{Raghavachari:89}
and with inclusion of higher excitations it is possible to smoothly converge to the exact answer. 
\cite{tajti_heat:_2004,bomble_high-accuracy_2006,harding_high-accuracy_2008}
It does not mean, however, that the coupled cluster hierarchy is the most efficient way to approach the FCI limit.
On the contrary, results from the distinguishable cluster singles and doubles (DCSD) \cite{kats_dc_2013,kats_dcsd_2014,kats_accurate_2015,kats_distinguishable_2016,kats_improving_2018}
 and other methods \cite{Meyer:71,paldus_approximate_1984,piecuch_solution_1991,piecuch_approximate_1996,kowalski_method_2000,
bartlett_addition_2006,nooijen_orbital_2006, Neese:09, huntington_pccsd:_2010,robinson_approximate_2011,
huntington_accurate_2012,paldus_externally_2016,black_statistical_2018}
are usually much better than the coupled cluster singles and doubles (CCSD) results. 
Additionally, the DCSD equations can be implemented more efficiently than CCSD by employing
density fitting or other integral decomposition techniques. \cite{kats_sparse_2013}
Recently, two publications have demonstrated that DCSD can be calculated exactly with 
the ${\cal O}(N^5)$ computational scaling by changing the representation,
whereas for CCSD still the ${\cal O}(N^6)$ scaling remains. \cite{hummel_low_2017,mardirossian_lowering_2018} 
This raises the question whether there exists a distinguishable cluster hierarchy which would  
produce more accurate results than the corresponding coupled cluster counterparts also for higher excitations,
with a lower nominal scaling. 

The distinguishable cluster (DC) doubles equations can be formulated using a screened Coulomb formalism, 
which can be used to introduce a perturbative triples correction to DCSD 
that was shown to be able to improve the relative energies from DCSD. \cite{kats_distinguishable_2016}
Here we will follow a different approach, more closely related to the original formulation of the 
DC approximation: we use the coupled cluster amplitude equations, remove (some) exchange
terms and restore the exactness for $n$ electrons. 
Moreover, in this contribution we will investigate the applicability of the DC approximation to the 
higher order excitation, and therefore we leave the doubles CCSDT amplitude equations untouched. 
As the result, we do not aim here at reproducing the accuracy of DCSD in the strongly correlated regime,
but rather at seeing what gain in the accuracy from the DC approximation can be expected for higher excitations.

\section{Theory}
\subsection{DC-CCSDT}
\label{sec:dc-ccsdt}
\begin{figure}
\centering
%
\includegraphics{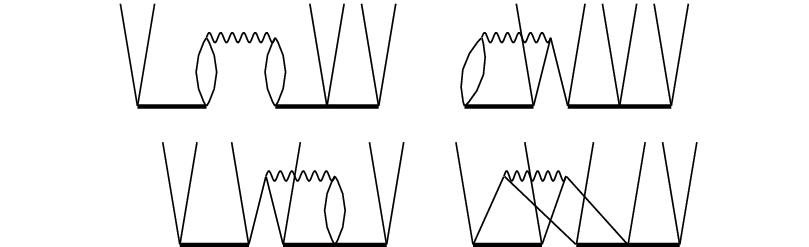}
\caption{$\op T_2 \op T_3$ skeleton diagrams in CCSDT equations.
}
\label{fig:t2t3}
\end{figure}

As in the original DC formulation, we consider singles to be responsible for orbital relaxation only,
and therefore focus our attention only to doubles and triples amplitudes. 
The CCSDT triple amplitudes equations using spin-orbitals can be found e.g. in Ref.~\onlinecite{shavitt_many-body_2009}.
Here we repeat the most relevant part of the CCSDT triples equations which contain doubles and triples 
connected by the two-electron repulsion integrals.
\begin{eqnarray}
R^{ijk}_{abc}&&\leftarrow \Perm{i}{jk}\Perm{a}{bc} \Big\{ (ld|me)T^{il}_{ad} T^{mjk}_{ebc}
-(ld|me)T^{il}_{ae} T^{mjk}_{dbc}\Big\} \nl
-\Perm{i}{jk}(ld|me)T^{li}_{de} T^{mjk}_{abc} -\Perm{a}{bc}(ld|me)T^{lm}_{da}T^{ijk}_{ebc}\nl
-\Perm{k}{ij}\Perm{a}{bc}(ld|me)T^{ij}_{ad}T^{lmk}_{bec} \\ &&
-\Perm{i}{jk}\Perm{c}{ab}(ld|me) T^{il}_{ab}T^{jmk}_{dec}\nl
+\half\Perm{k}{ij}(ld|me) T^{ij}_{de} T^{lmk}_{abc}+\half\Perm{c}{ab}(ld|me)T^{lm}_{ab}T^{ijk}_{dec}\nonumber
\label{eq:t2t3}
\end{eqnarray}
where the permutation operator $\Perm{p}{qr}$ antisymmetrizes a tensor as
\begin{eqnarray}
\Perm{p}{qr} X_{pqr}&& = X_{pqr} - X_{qpr} - X_{rqp}.
\end{eqnarray}
Here and in the following indices $i,j,k,l,m$ denote occupied spin-orbitals, $a,b,c,d,e$ virtual,
and $p,q,r,s$ are general spin-orbital indices, and we assume the Einstein convention for the
repeated indices.
$T^{ij}_{ab}$, $T^{ijk}_{abc}$, and $(ld|me)$ are the doubles and triples amplitudes, and the
two-electron integrals in the chemical (Mulliken) notation, respectively. 
Note that the integrals here are not antisymmetrized; the amplitudes, however, still possess the proper 
permutational antisymmetry ensured by the permutation operators in \eq{eq:t2t3}.
The skeleton diagrams corresponding to these terms are shown in \fig{fig:t2t3}.

In the distinguishable cluster approximation we remove diagrams, which correspond to 
intercluster exchange, in the local particle-hole symmetric fashion, \cite{kats_distinguishable_2016,kats_particlehole_2018} 
i.e., the last diagram in \fig{fig:t2t3}.
By restricting the $i,j,k$ indices in \eq{eq:t2t3} to $i=1,j=2,k=3$ (which is the only possibility 
if one has only three occupied spin-orbitals and the usual restriction $i\lt j\lt k$) one can show that the terms 
originating from this exchange skeleton diagram (the second, seventh, and eighth terms in \eq{eq:t2t3}) cancel 
with one half of the terms originating from the second and third skeleton diagram in \fig{fig:t2t3}
(in particular, the fourth and fifth, third, and sixth terms in \eq{eq:t2t3}, respectively)
in the case of three electrons. This cancellation is unique (as long as we restrict ourselves to 
particle-hole symmetric methods).
Therefore, removing the exchange diagram and putting a factor one half in front of the two other 
diagrams yields a local particle-hole symmetric method that is exact for three electrons. This method will 
be denoted as DC-CCSDT in following. 
Note that the fourth and the fifth terms in \eq{eq:t2t3} are exactly equal in the case of three electrons,
and therefore in principle one can take any linear combination of them to cancel the second term.
However, we chose the equally weighted linear combination which corresponds to a 
local-particle-hole symmetric equations. 

By applying the same analysis as in Ref.~\onlinecite{mardirossian_lowering_2018}, it is straightforward to demonstrate that this method
has a nominal scaling ${\cal O}(N^7)$ with the systems size $N$, as opposed to ${\cal O}(N^8)$ scaling of CCSDT. 
Let us focus at the first and the second term from \eq{eq:t2t3} to show that the first term can
be calculated with ${\cal O}(N^7)$ scaling by going to the real-space representation, whilst 
the second term remains ${\cal O}(N^8)$ scaling. 
In \eq{eq:coul} a route is sketched how to calculate the first term of \eq{eq:t2t3} with the ${\cal O}(N^7)$
scaling. 
\begin{eqnarray}
\label{eq:coul}
&&(ld|me)T^{il}_{ad} T^{mjk}_{ebc}=\\&&
\int d\mat r_1 d\mat r_2\phi_l^*(\mat r_1) \phi_d(\mat r_1)
\frac{1}{|\mat r_1 - \mat r_2|} \phi_m^*(\mat r_2) \phi_e(\mat r_2)
T^{il}_{ad} T^{mjk}_{ebc}\nl
=\int d\mat r_2 T^{jk}_{bc}(\mat r_2)\int d\mat r_1 T^{i}_{a}(\mat r_1) \frac{1}{|\mat r_1 - \mat r_2|}\nl
{\rm with}\nl
T^{jk}_{bc}(\mat r_2)=\phi_m^*(\mat r_2) \phi_e(\mat r_2) T^{mjk}_{ebc}\nl
T^{i}_{a}(\mat r_1)=\phi_l^*(\mat r_1) \phi_d(\mat r_1) T^{il}_{ad}
\nonumber
\end{eqnarray}
The same (but more practical) scaling reduction for the $\op T_2 \op T_3$ DC-CCSDT terms 
can be achieved by employing the density fitting approximation, 
as has been demonstrated previously for DCSD. \cite{kats_sparse_2013}
However, for some other terms in the triples equations, e.g., the $\op T_3$-ladder diagrams,
the real-space representation \cite{mardirossian_lowering_2018} or the tensor hypercontraction \cite{parrish_communication:_2014,hummel_low_2017}
have to be used to reduce the scaling to ${\cal O}(N^7)$ in the same way as for the $\op T_2$-ladder diagrams
in the doubles amplitude equations.

For the exchange term on the other hand this simplification would not work, since no 
reduction of the dimensionality with respect to the system size of the amplitudes can be achieved by 
going to the real space. 


\subsection{Connection to the screened Coulomb formalism}
\begin{figure}
\centering
\includegraphics{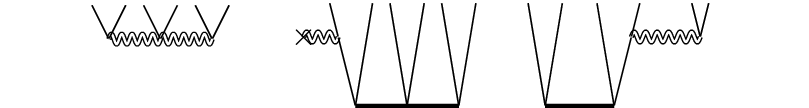}
\caption{Origin of $\op T_2 \op T_3$ diagrams in screened DC-CCSDT equations.
}
\label{fig:screen}
\end{figure}

In this section we briefly discuss a connection to the screened Coulomb formalism. 
In the screened Coulomb formalism from Ref.~\onlinecite{kats_distinguishable_2016} the screening 
is done using a direct-random-phase-approximation 
dressing of the fluctuation densities. The internally dressed Fock matrices have been obtained by
switching to a local-particle-hole formulation of the Fock matrix. The same can be done to obtain
the $\op T_2 \op T_3$ terms. We have to extend our dressing by introducing a three-body dressed 
integrals,
\begin{eqnarray}
(pq|ai|bj)&&=(pq|kc)T^{kij}_{cab},
\label{eq:threebody}
\end{eqnarray}
which contribute to effective two-body integrals after an exchange-Fock-like internal contraction 
\begin{eqnarray}
\widetilde{(aq|bj)}\leftarrow-\half \delta_{pi}(pq|ai|bj)&&=-\half(iq|kc)T^{kij}_{cab},\nonumber\\
\widetilde{(ip|bj)}\leftarrow-\half \delta_{qa}(pq|ai|bj)&&=-\half(pa|kc)T^{kij}_{cab}.
\end{eqnarray}
With this screening all the $\op T_2 \op T_3$ terms in DC-CCSDT can be obtained trough 
screening of linear terms in the triples equations, see \fig{fig:screen}.
The first diagram in \fig{fig:t2t3} contributes to the first diagram in \fig{fig:screen}, and the 
second and third diagrams in \fig{fig:screen} contain the second and third diagrams from \fig{fig:t2t3}
scaled by a factor one half.

\section{Test calculations}
\label{sec:results}

An initial test implementation of the DC-CCSDT equations from \sect{sec:dc-ccsdt} was obtained by an automated term generation approach, as implemented in the GeCCo program.\cite{gecco}  
The CCSDT, CCSDT(Q), and CCSDTQ results were obtained using either the GeCCo or MRCC\cite{mrcc} implementations,
and Molpro \cite{molpro} was used for the Hartree-Fock (HF) calculations and for generation of the integrals.
We employed the cc-pVDZ basis in all calculations apart from the beryllium dimer calculations, for which the aug-cc-pVTZ basis was used, 
together with the frozen core approximation. 

Our present implementation is not applicable to molecular systems with unpaired electrons, 
therefore we tested its accuracy for three-electron systems by calculating the BeH$_2$ molecule with one of the hydrogen atoms at a distance of
{100~\AA} from the beryllium atom 
(the HBeH angle was set to 104 degree and the other BeH bond length to {1.1~\AA}). 
The results are presented in \tab{tab:beh2}.
As expected the new method is very accurate for this quasi-three-electron system. 
We also see that it is much more accurate than the CCSDT method for breaking the BeH single bond.
The error is comparable to the UCCSDT method, where the discrepancy compared to the FCI results comes from the
difference in core orbitals between UHF and closed-shell RHF employed in UCCSDT and the rest, respectively.

\begin{table}[htbp]                                                                                 
\centering                                                                                          
\caption{Results for a BeH$_2$ molecular system with one of the BeH bonds stretched to {100~\AA}.
}
\begin{tabular}{lcc}                                                                               
\hline                                                                                              
          & Energy, $E_h$& Error, $\mu E_h$ \\                                                                
\hline                                                                                              
FCI       & -15.662290 & 0  \\                                                               
CCSDT     & -15.662155 & 135\\                                                               
UCCSDT$^a$& -15.662305 & -15\\                                                               
DC-CCSDT  & -15.662271 & ~19 \\                                                               
\hline                                                                                              
\end{tabular}%
\label{tab:beh2}                                                                                     

$^a$ Calculated using the MRCC code.\cite{mrcc}
\end{table}

\begin{figure}
\centering
\includegraphics[width=0.9\linewidth]{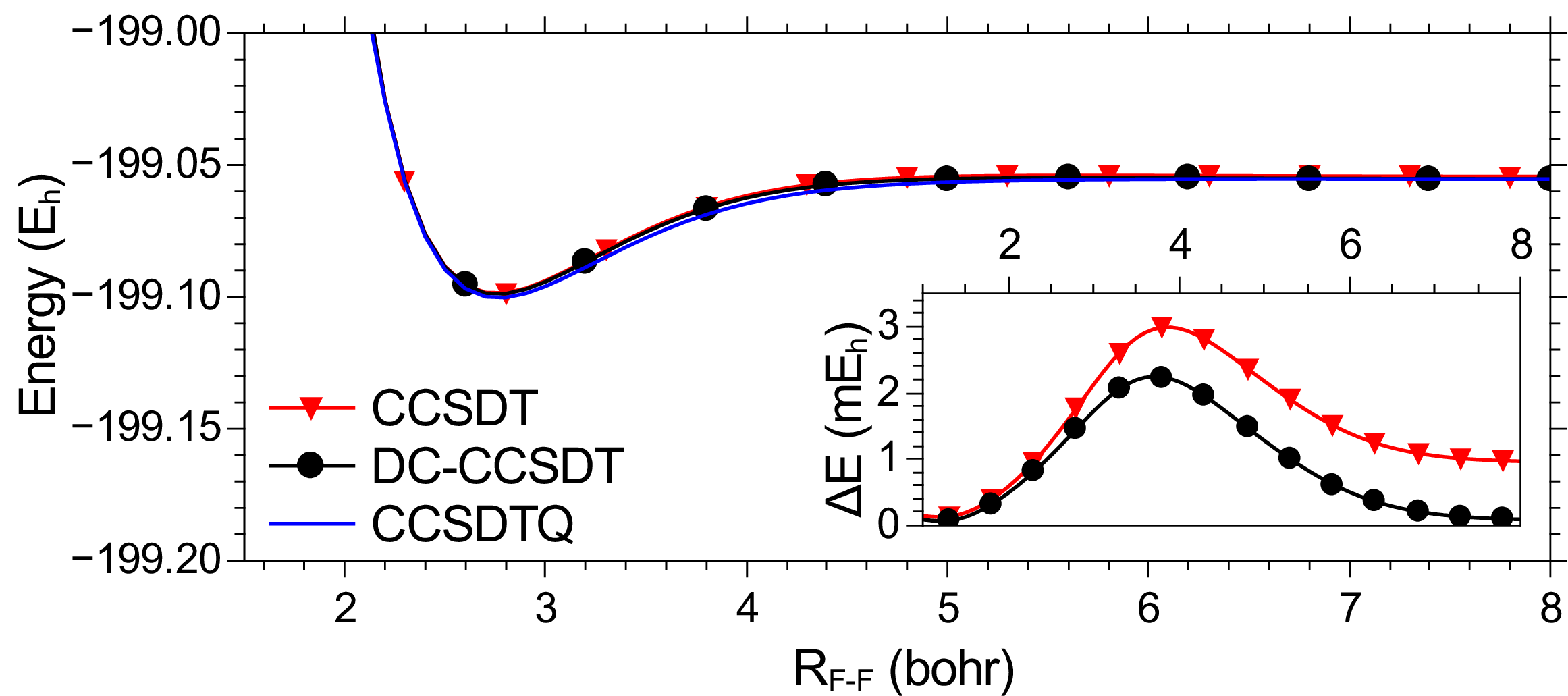}
\caption{Potential energy curve of the F$_2$ molecule from CCSDT, CCSDTQ, and DC-CCSDT.
The difference plot is given with respect to CCSDTQ.
}
\label{fig:f2}
\end{figure}

To verify the higher accuracy of DC-CCSDT in breaking single bonds compared to the normal CCSDT method
we calculated the potential energy curve of the F$_2$ molecule. The plot can be found in \fig{fig:f2}. 
While the error of CCSDT remains at 1 m$E_h$ in the asymptotic region, DC-CCSDT approaches the reference result.
At 8 bohr separation the energy difference between the F$_2$ molecular system and two separate fluorine atoms 
is 0.1 m$E_h$ for CCSDTQ and 0.9 m$E_h$ for CCSDT, which verifies the CCSDT issues in breaking single bonds.

\begin{figure}
\centering
\includegraphics[width=0.9\linewidth]{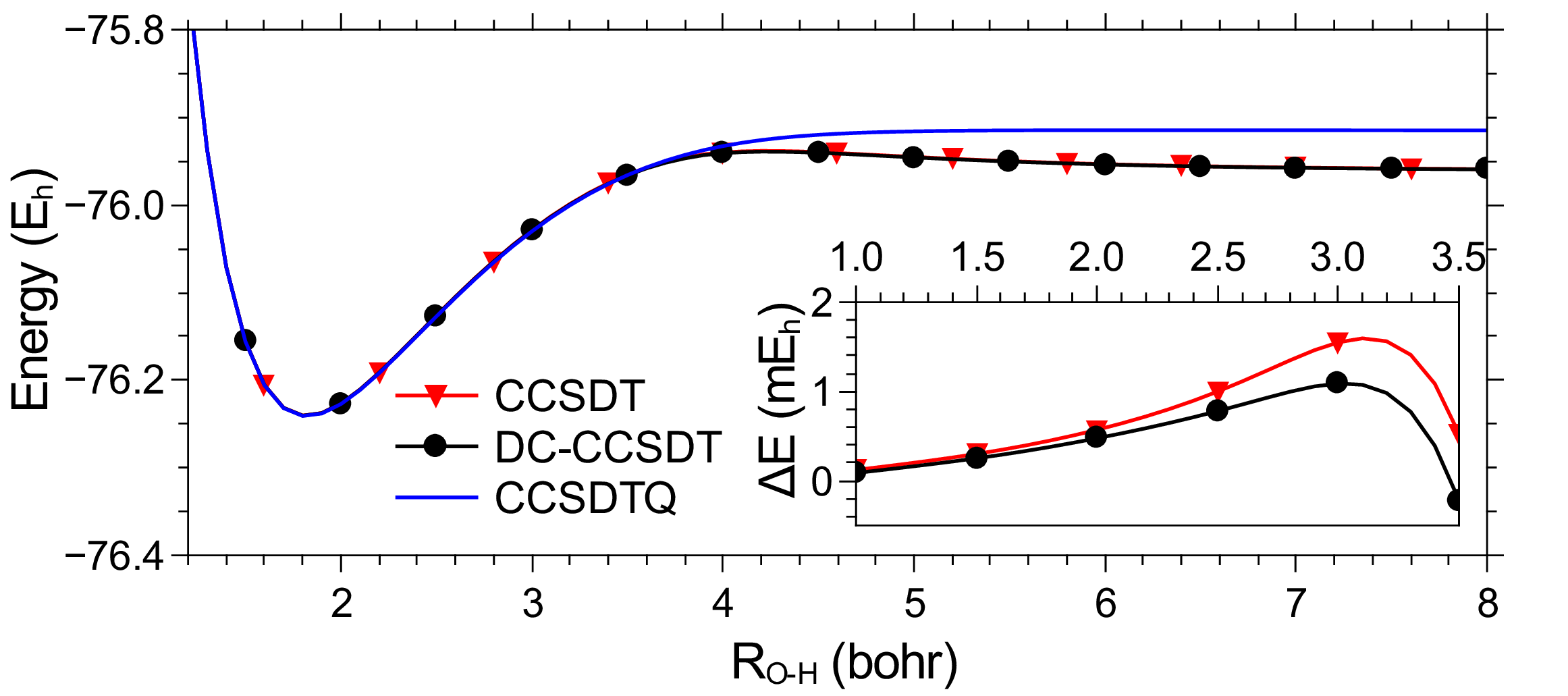}
\caption{Potential energy curve of the simultaneous dissociation of two hydrogen atoms from a water molecule 
from CCSDT, CCSDTQ, and DC-CCSDT. HOH angle was set to 107.6 degree. 
The difference plot is given with respect to CCSDTQ.
}
\label{fig:h2o}
\end{figure}

As expected, the DC-CCSDT method shows problems for the simultaneous breaking of two bonds, since we have reintroduced
all CCSD quadratic doubles terms in the doubles amplitude equations. Indeed, DC-CCSDT  
fails for the simultaneous dissociation of two hydrogen atoms from the water molecule, \fig{fig:h2o}. 
However, the curve in the intermediate region is noticeably closer to the CCSDTQ reference than the CCSDT 
curve, which demonstrates again the enhanced accuracy of the DC-CCSDT. 

\begin{figure}
\centering
\includegraphics[width=0.9\linewidth]{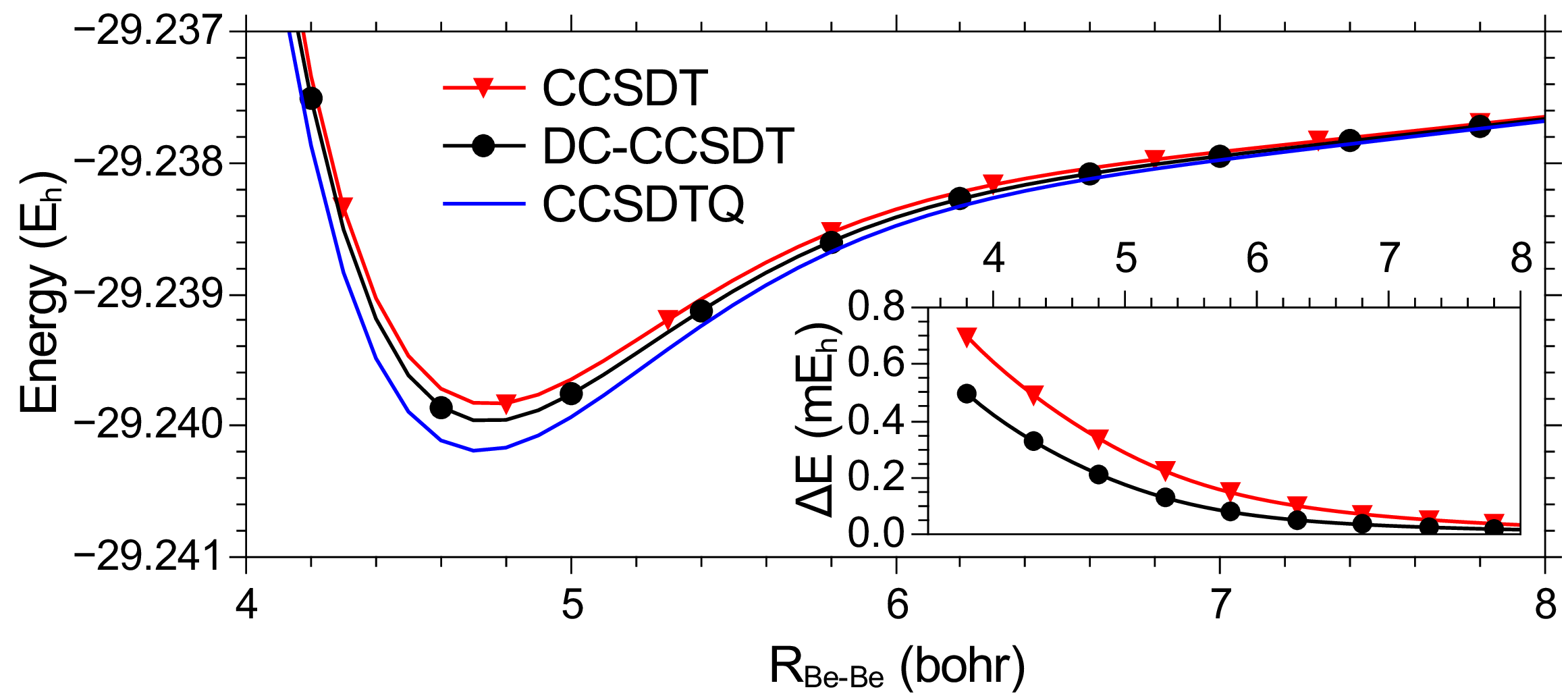}
\caption{Potential energy curve of the beryllium dimer from CCSDT, CCSDTQ, and DC-CCSDT.
The difference plot is given with respect to CCSDTQ.
}
\label{fig:be2}
\end{figure}

The beryllium dimer is another molecular system which is known to require high level theories for
a quantitatively accurate description. \cite{lesiuk_ab_2019} 
We used the aug-cc-pVTZ basis set and compared the accuracy 
of CCSDT and DC-CCSDT to CCSDTQ (which in the frozen core case, as employed here, is equivalent to FCI).
The potential energy curves together with an error curve demonstrating the discrepancy of CCSDT and DC-CCSDT
curves to the CCSDTQ results are shown in \fig{fig:be2}. 
Clearly the distinguishable cluster approximation improves again upon the CCSDT results. For example,
around the equilibrium distance (which in these calculations is around 4.75 bohr) the error of DC-CCSDT
is approximately 0.22 m$E_h$, while that of CCSDT is approximately 0.35 m$E_h$.

\begin{figure}
\centering
\includegraphics[width=0.9\linewidth]{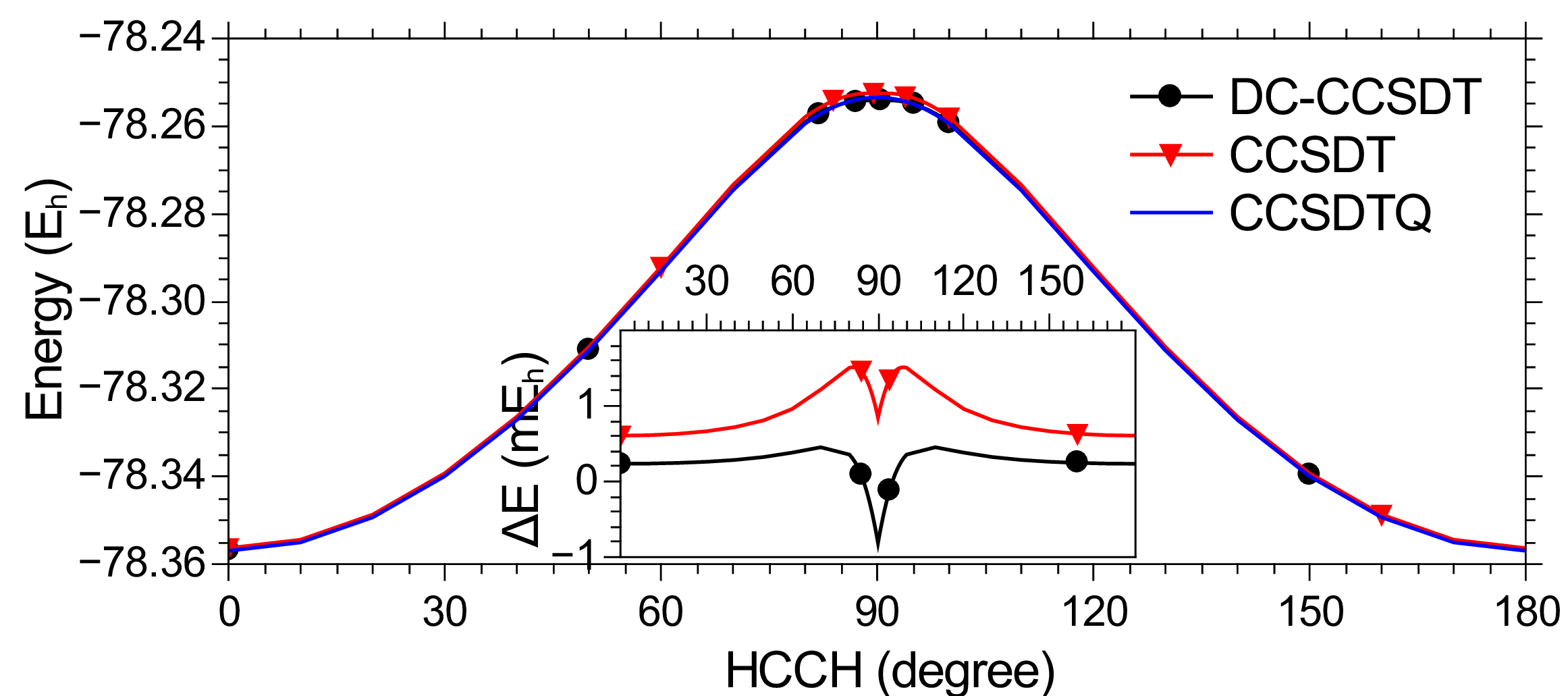}
\caption{Potential energy curve of twisting the ethylene molecule from CCSDT, CCSDTQ, and DC-CCSDT.
The difference plot is given with respect to CCSDTQ.
}
\label{fig:c2h4}
\end{figure}

Twisting the ethylene molecule is one of the problems where CCSDT is considered to be very accurate.
Here we compared the accuracy of CCSDT and DC-CCSDT versus CCSDTQ results. 
We varied the HCCH angle from 0 to 90 degrees and fixed each of C-CH2 blocks to be planar. 
The remaining degrees of freedom ($r_{\rm CC}$, $r_{\rm CH}$, and HCH angle) were optimised 
at the DCSD level, which has been demonstrated to be very accurate for this system. \cite{kats_accurate_2015} 
The plots of the potential energy curve and the errors of CCSDT and DC-CCSDT compared to CCSDTQ 
can be found in \fig{fig:c2h4}.
Although the potential energy curves look very similar, the errors of CCSDT reach over 1.5 m$E_h$ (nearly 1 kcal/mol)
for geometries around 80 degrees twisting. The DC-CCSDT errors in this region are comfortably low, 
less than 0.5 m$E_h$ ( $\lt0.3$ kcal/mol), and even the largest error encountered at 90 degrees is only 0.8 m$E_h$ (0.5 kcal/mol).

The accuracy of relative energies is much more important for the applications than the accuracy of
absolute energies. In order to investigate it, we have calculated 36 reaction energies involving 
closed-shell molecules. \cite{huntington_accurate_2012,kats_improving_2018,kats_orbital-optimized_2019} 
Mean absolute deviations (MAD), root-mean squared deviations (RMSD), and maximal deviations (MaxD)
with respect to the CCSDT(Q) results
can be found in \tab{tab:react}. 
The MAD and RMSD improve only slightly by going from perturbative triples to full CCSDT triples 
(improvement in the range $10-15\%$)
However, DC-CCSDT reaction energies are much closer to CCSDT(Q) results and demonstrate an 
improvement of around one third upon CCSD(T) results.
If one removes the reaction which shows the largest deviation for all three methods considered here
(reaction 27 from Ref.~\onlinecite{huntington_accurate_2012}, which is also the most exothermic reaction 
in the set, with over 100 kcal/mol larger reaction energy than the next one), 
the results become even more striking: the MAD of
DC-CCSDT goes below 0.1 kcal/mol, and the improvement upon CCSD(T) results increases to $40-45\%$.
The CCSDT improvement on the other hand remains around $10\%$.
\begin{table}
\centering                                                                                          
\caption{ Statistical analysis of the absolute deviations of CCSD(T), CCSDT, and DC-CCSDT 
from CCSDT(Q) results (in kcal/mol).}
\begin{tabular}{lccc}                                                                               
\hline                                                                                              
                  & MAD   & RMSD  & MaxD \\                                                                
\hline                                                                                              
CCSD(T)          & 0.178 & 0.263 & 0.700 \\                                                               
CCSDT            & 0.160 & 0.224 & 0.534 \\                                                               
DC-CCSDT         & 0.114 & 0.176 & 0.685 \\                                                               
                 & \multicolumn{3}{c}{without reaction 27 from Ref.~\onlinecite{huntington_accurate_2012}}\\
CCSD(T)          & 0.163 & 0.239 & 0.698 \\                                                               
CCSDT            & 0.150 & 0.208 & 0.529 \\                                                               
DC-CCSDT         & 0.098 & 0.135 & 0.334 \\                                                               
\hline                                                                                              
\end{tabular}%
\label{tab:react}                                                                                     
\end{table}

\section{Conclusions}
We applied the distinguishable cluster approximation to the $\op T_2 \op T_3$ terms of the
CCSDT triples amplitude equations, and demonstrated that the DC-CCSDT method is superior to
CCSDT in all cases we have investigated so far. Although it lacks the ability of the DCSD method 
to qualitatively correctly break multiple bonds, the new method is noticeably more accurate than 
CCSDT in breaking one or two bonds. For example, it seems to dissociate the F$_2$ molecule to the
correct limit, whilst CCSDT has some remaining error even at the interatomic distance of 8 bohr. 
The reaction energies are also much more accurate than from CCSDT, which barely improves upon CCSD(T).
All the benchmarks clearly indicate that the improvement upon the CC hierarchy seen on the DCSD level can also 
be achieved for higher order methods.

Additionally this work demonstrates that a size-extensive orbital-invariant method, 
exact for three electrons, can be formulated with a nominal computational scaling of ${\cal O}(N^7)$ 
with respect to the system size. 

Further work is required to make it applicable to strong electron correlation in the similar way as
DCSD, and is the focus of our ongoing research.

\begin{acknowledgements}
DK acknowledges financial support from the Max-Planck Society.
\end{acknowledgements}

%
\end{document}